\documentclass[twocolumn,trackchanges]{aastex631}
\usepackage{natbib}
\usepackage{epsfig}
\usepackage{graphicx}
\usepackage{subfigure}
\usepackage{float}
\usepackage{amsmath}
\usepackage{color}
\usepackage{amssymb}
\usepackage{apjfonts}

\newcommand{\he}{\mbox{He {\sc ii}}\;}

\newcommand{\PMO}{Purple Mountain Observatory, Chinese Academy of Sciences, Nanjing 210023, China}
\newcommand{\USTC}{School of Astronomy and Space Sciences, University of Science and Technology of China, Hefei 230026, China}

\begin{document}

\title{Forecasts for Helium Reionization Detection with Fast Radio Bursts in the Era of Square Kilometre Array}

\correspondingauthor{Jun-Jie Wei}
\email{jjwei@pmo.ac.cn}

\author[0000-0003-0162-2488]{Jun-Jie Wei}
\affiliation{\PMO}
\affiliation{\USTC}

\author{Chong-Yu Gao}
\affiliation{\PMO}
\affiliation{\USTC}

\begin{abstract}
The observed dispersion measures (DMs) of fast radio bursts (FRBs) are a good indicator of the amount of ionized material
along the propagation paths. In this work, we present a forecast of \he reionization detection using the DM and 
redshift measurements of FRBs from the upcoming Square Kilometre Array (SKA). Assuming a model of the Universe 
in which \he reionization occurred at a specific redshift $z_{\rm re}$, we analyze what extent the signal-to-noise
ratio ($\mathrm{S/N}$) for the detection of the amplitude of reionization can be achieved in the era of SKA. 
Using $10^{6}$ mock FRB data from a one-year observation of the second phase of SKA, we find that the $\mathrm{S/N}$
for detecting \he reionization can approach the $32-50\sigma$ level and the uncertainty on the reionization redshift
can be constrained to be $\sigma(z_{\rm re})\approx 0.022-0.031$, depending on the assumed FRB redshift distribution.
This is the first quantitative analysis on the detection significance of \he reionization in the SKA era.
We also examine the influence of different fiducial $z_{\rm re}$ values, finding that this effect has a modest impact 
on the forecasts. Our results demonstrate the potentially remarkable capability of SKA-era FRBs in constraining
the epoch of \he reionization.
\end{abstract}


\section{Introduction}
\label{sec:intro}
Reionization is regarded as the latest phase transition in the Universe, transforming the baryonic gas
from a neutral condition into an ionized state \citep{2001PhR...349..125B}. Understanding how and when 
the cosmic reionization happened is one of the important research topics in modern cosmology. Reionization 
usually refers to the process by which neutral hydrogen and helium atoms lost their first outer 
electrons at redshifts $z\ga 6$ (e.g., \citealt{2002AJ....123.1247F,2013ASSL..396...45Z,2018ApJ...858...54S}).
During the epoch of hydrogen reionization, the high ionization potential ($\mathrm{54.4\,eV}$)
of singly ionized helium (\mbox{He {\sc ii}}) kept its inner electron from being further ionized.
The reionization of \mbox{He {\sc ii}}, caused by sufficient strong radiation from quasars and stars, 
is generally believed to occur at $z\sim3$. The strongest evidence of \he reionization is obtained from 
observational signatures of the \he Ly$\alpha$ forest in the spectra of quasars at $z\sim3$ 
\citep{2009ApJ...690.1181S,2012AJ....143..100S,2024MNRAS.532..841B}. Other evidence is the temperature change of 
the intergalactic medium (IGM; \citealt{2008ApJ...681....1F,2008ApJ...682...14F,2009ApJ...694..842M}). 
The relatively lower redshift of \he reionization enables it more likely to be detected by 
next generation surveys.

Fast radio bursts (FRBs) are bright millisecond radio pulses predominantly originate at 
cosmological distances, the study of which is a hot spot in contemporary astrophysics 
\citep{2007Sci...318..777L,2019ARA&A..57..417C,2019A&ARv..27....4P,2022A&ARv..30....2P,2021SCPMA..6449501X,2023RvMP...95c5005Z}.
Their all-sky rate is estimated to be as high as $\sim10^{4}\,\mathrm{events\,day^{-1}}$ 
\citep{2013Sci...341...53T}. A direct observable of FRBs is the dispersion measure (DM) 
that encodes the integrated column density of ionized plasma along the line of sight.
The observed DM is thus a good indicator of the amount of ionized baryons along the 
propagation path between the FRB source and the observer. Moreover, some FRBs with extremely 
high DM measurements are expected to be detected up to $z\ga 6$ by future large-aperture 
telescopes \citep{2017ApJ...846L..27F,2018ApJ...867L..21Z,2020MNRAS.497.4107H}. Thanks to 
the short pulse nature, the high event rate, and the cosmological origin, FRBs are suitable for 
being used as a potentially new probe for the ionized IGM. Several studies have proposed that 
future DM measurements of high-$z$ FRBs could be used to constrain the reionization history 
of hydrogen (e.g., \citealt{2014ApJ...783L..35D,2021MNRAS.502.5134B,2021JCAP...05..050D,2021MNRAS.502.2346H,
2021MNRAS.505.2195P,2022ApJ...933...57H,2024arXiv240805722M,2024arXiv240903255S}) and helium 
(e.g., \citealt{2014ApJ...783L..35D,2014ApJ...797...71Z,2019MNRAS.485.2281C,
2020PhRvD.101j3019L,2021PhRvD.103j3526B,2021NewA...8901627L,2022Univ....8..317J}).
Since the current sample size and redshift range of localized FRBs are small, all such studies 
were performed by building mock FRB data sets.

As a next-generation radio astronomy-driven Big Data facility \citep{2015aska.confE..55M}, 
the Square Kilometre Array (SKA) will definitely have very powerful capability of detecting
and localizing high-$z$ FRBs \citep{2017ApJ...846L..27F,2020MNRAS.497.4107H}, which will 
offer a much larger population of FRBs for exploring the epoch of reionization. Therefore, 
it is natural to ask what level of reionization constraints can be achieved using future 
FRB data in the era of SKA. In view of the fact that \he reionization happened at a 
comparatively lower redshift and is more accessible to measurements, here we investigate 
the ability of the SKA-era FRB observation to detect \he reionization through Monte Carlo 
simulations.

But though FRBs have previously been proposed to constrain the epoch of \he reionization
\citep{2014ApJ...783L..35D,2014ApJ...797...71Z,2019MNRAS.485.2281C,
2020PhRvD.101j3019L,2021PhRvD.103j3526B,2021NewA...8901627L,2022Univ....8..317J},
it has not yet been quantitatively analyzed what extent the nature of \he reionization 
can be explored in the SKA era---our primary focus in this paper. Previously, 
\cite{2014ApJ...783L..35D} and \cite{2014ApJ...797...71Z} first proposed the potential
application of FRBs in probing \he reionization, \cite{2019MNRAS.485.2281C} estimated 
the number of FRBs required to distinguish between a sudden \he reionization that occurred
at $z=3$ or $6$ by using a two-sample Kolmogorov-Smirnoff test, \cite{2020PhRvD.101j3019L}
assessed the ability of future high-$z$ FRB samples to detect \he reionization, 
\cite{2021PhRvD.103j3526B} showed that the statistical ensemble of the DM distribution 
of FRBs prove useful probes of \he reionization even with limited redshift information,
\cite{2021NewA...8901627L} presented a cosmology-based model on determining
the \he reionization redshift in range of $z = 3$ to $4$ by detecting 
the DMs of distant FRBs, and \cite{2022Univ....8..317J} attempted to use the DM--$z$
measurements of FRBs to derive the helium abundance in the Universe. In this work,
we build on the techniques outlined by these authors by performing a forecast 
of the detection significance of \he reionization with the SKA-era FRB observation.

The rest of this paper is organized as follows. In Section~\ref{sec:DM}, we show the imprint 
of the occurrence of \he reionization on the redshift function of the IGM portion of the DM, 
i.e., $\mathrm{DM_{IGM}}(z)$, and describe the definition of a signal-to-noise ratio ($\mathrm{S/N}$) 
for the detection of \he reionization. The expected detection rates of FRBs by SKA, the 
simulation method, and the relevant analysis results are presented in Section~\ref{sec:simulation}. 
Finally, we give a brief summary and discussion in Section~\ref{sec:sum}. Throughout this paper, 
we assume $\Lambda$CDM cosmology and adopt the cosmological parameters of $H_{0}=67.4$ km $\rm s^{-1}$ $\rm Mpc^{-1}$, 
$\Omega_{\rm m}=0.315$, $\Omega_{b}=0.0493$, and $\Omega_{\Lambda}=0.685$ \citep{2020AA...641A...6P}.

\section{Impact of Helium Reionization on Dispersion Measure}
\label{sec:DM}
For an extragalactic FRB propagating through the IGM, the cosmic DM in the observer frame is related to 
the free electron density ($n_{\rm e}$) along the line of sight:
\begin{equation}
\mathrm{DM_{IGM}}(z)=\int_{0}^{z}\frac{n_{\rm e}(z')}{1+z'}\mathrm{d}l'\;.
\end{equation}
In a flat $\Lambda$CDM cosmology, the relation between proper length element $\mathrm{d}l'$ and redshift $z'$ 
is given by
\begin{equation}
\mathrm{d}l'=\frac{1}{1+z'}\frac{c}{H_0}\frac{\mathrm{d}z'}{\sqrt{\Omega_{\rm m}(1+z')^3+\Omega_{\Lambda}}}\;,
\end{equation}
where $c$ is the speed of light, $H_0$ is the Hubble constant, $\Omega_{\rm m}$ is the present-day matter density,
and $\Omega_{\Lambda}=1-\Omega_{\rm m}$ is the vacuum energy density. The number density of free electrons
can be expressed as \citep{2014ApJ...783L..35D}
\begin{equation}
n_{\rm e}(z)=\frac{\rho_{c,0}\Omega_{b}f_{\rm IGM}X_{e}(z)}{m_p}(1+z)^3\;,
\end{equation}
where $\rho_{c,0}\equiv3H_{0}^{2}/8\pi G$ is the critical density of the Universe, $m_p$ is the proton mass,
$\Omega_{b}$ is the baryon density parameter, and $f_{\rm IGM}\simeq0.83$ is the mass fraction of baryons 
in the IGM \citep{1998ApJ...503..518F}. The electron fraction is  
\begin{equation}
X_{e}(z)=Y_{\rm H} X_{e,\mathrm{H\,II}}(z)+\frac{1}{4}Y_{\rm He}\left[X_{e,\mathrm{He\,II}}(z)+2X_{e,\mathrm{He\,III}}(z)\right]\;,
\end{equation}
where $Y_{\rm H}=1-Y_{\rm He}$ and $Y_{\rm He}$ are the hydrogen and helium abundances in the Universe, 
respectively, and $X_{e,\mathrm{H}}(z)$ and $X_{e,\mathrm{He}}(z)$ represent the ionization mass fractions 
of hydrogen and helium, respectively. The singly and doubly ionized elements are denoted by subscripts 
``II'' and ``III'', respectively. At a given redshift, the measured value of $\mathrm{DM_{IGM}}$ would 
vary along different lines of sight due to the large density fluctuations in IGM. In theory, the averaged 
$\mathrm{DM_{IGM}}$ over different lines of sight can be estimated by \citep{2003ApJ...598L..79I,2004MNRAS.348..999I}
\begin{equation}
\mathrm{DM_{IGM}}(z)=\frac{3 c H_{0} \Omega_{b} f_{\mathrm{IGM}}}{8 \pi G m_{p}} \int_{0}^{z} \frac{(1+z^{\prime}) X_{e}(z^{\prime})}{\sqrt{\Omega_{\rm m}(1+z^{\prime})^3+\Omega_{\Lambda}}} \mathrm{d}z^{\prime}\;.
\label{eq:DMIGM}
\end{equation}

\begin{figure}
\begin{center}
\vskip-0.3in
\includegraphics[width=0.45\textwidth]{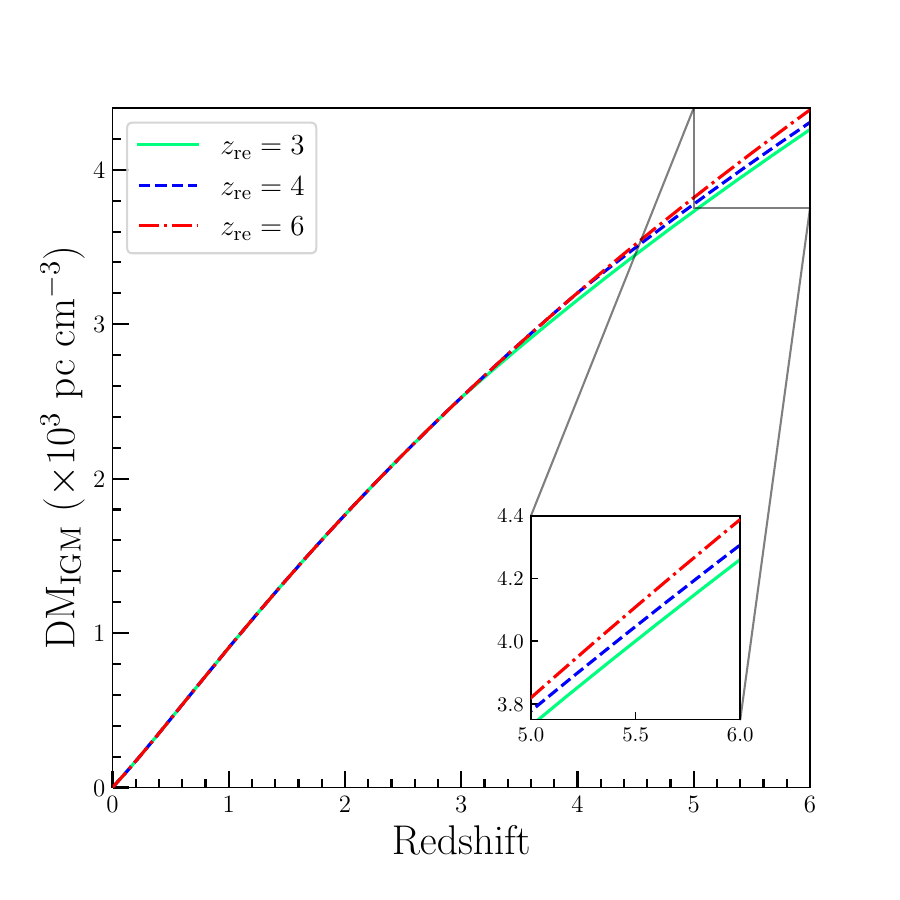}
\vskip-0.1in
\caption{Sensitivity of the cosmic dispersion measure $\mathrm{DM_{IGM}}(z)$ function to \he reionization. 
A sudden \he reionization is assumed to occur at redshift $z_{\rm re}=3,\,4$, and $6$, respectively. 
The inset illustrates a zoomed-in view of the separation among different model curves.}
\label{fig1}
\vskip-0.2in
\end{center}
\end{figure}

At the epoch of hydrogen reionization, both neutral hydrogen and helium are singly ionized by ionizing photons
emitted from star-forming galaxies, $X_{e,\mathrm{He\,II}}$ can thus be approximated as $X_{e,\mathrm{H\,II}}$
(e.g., \citealt{2021JCAP...05..050D}). We suppose that hydrogen is fully ionized at $z=6$ (i.e., 
$X_{e,\mathrm{H\,II}}=1$ and $X_{e,\mathrm{He\,II}}=1$) and that \he reionization occurs
suddenly at redshift $z_{\rm re}$ (i.e., $X_{e,\mathrm{He\,II}}=0$ and $X_{e,\mathrm{He\,III}}=1$). That is, 
considering the FRB sample up to redshifts $z<6$, one has $X_{e,\mathrm{H\,II}}=1$, $X_{e,\mathrm{He\,II}}=1$,
and $X_{e,\mathrm{He\,III}}=0$ before \he reionization, while afterward $X_{e,\mathrm{H\,II}}=1$, 
$X_{e,\mathrm{He\,II}}=0$, and $X_{e,\mathrm{He\,III}}=1$. Thus, $X_{e}(z>z_{\rm re})=1-3Y_{\rm He}/4=0.819$
and $X_{e}(z<z_{\rm re})=1-Y_{\rm He}/2=0.880$, where we adopt the helium abundance inferred from Planck observations:
$Y_{\rm He}=0.241$ \citep{2020AA...641A...6P}. The ionization fraction increases by $\sim7.4\%$ because of
\he reionization. That is, the sudden \he reionization would cause a change in $\mathrm{DM_{IGM}}$ 
(Equation~\ref{eq:DMIGM}).

It is interesting to explore whether we can detect such \he reionization in the DM--$z$ measurements of 
a high-$z$ FRB sample. Following \cite{2020PhRvD.101j3019L}, we define a simple $\mathrm{S/N}$ criterion for
the detection of \he reionization. In practice, we model the detection as \citep{2020PhRvD.101j3019L}
\begin{equation}
\mathrm{DM_{IGM}}(z)=\mathrm{DM_{IGM}^{high}}(z)+A_{\rm He}\left[\mathrm{DM_{IGM}^{\emph{z}_{re}}}(z)-\mathrm{DM_{IGM}^{high}}(z)\right]\;,
\label{eq:SNR}
\end{equation}
where the superscripts ``high'' and ``$z_{\rm re}$'' mean \he reionization occurs beyond the redshift limit 
of the sample and at $z=z_{\rm re}$, respectively. If $A_{\rm He}=0$, then \he reionization is not detected 
by the sample, while $A_{\rm He}=1$ corresponds to reionization occurring at $z_{\rm re}$. With $\sigma(A_{\rm He})$ 
representing the uncertainty of the determined $A_{\rm He}$, the detection $\mathrm{S/N}$ is thus $A_{\rm He}/\sigma(A_{\rm He})$, 
or simply $1/\sigma(A_{\rm He})$ since it gives $A_{\rm He}=1$ if the modeling is right.

As an illustration, the DM of the IGM as a function of the redshift $z$ is given in Figure~\ref{fig1}. 
To show the effect of a sharp \he reionization on $\mathrm{DM_{IGM}}(z)$, we plot three cases with 
reionization occurring at $z_{\rm re}=3,\,4,$ and $6$, respectively. It is obvious that the cosmic DM 
has a change in slope at the reionization redshift due to the extra influx of electron fraction from 
\he reionization. The key questions are thus whether this signature (i.e., identify $A_{\rm He}\neq0$ with 
significant $\mathrm{S/N}$) can be detected, and whether the redshift $z_{\rm re}$ at which reionization occurs
can be measured. The model curves look pretty close together, as shown in Figure~\ref{fig1}, but 
it should be underlined that the $\mathrm{S/N}$ for the detection of \he reionization would be enhanced by
having enough high-$z$ FRBs on either side of the reionization redshift. In the next section,
we will investigate the prospect of \he reionization detection with the SKA-era FRB observation.

\section{Forecasting with FRB mock}
\label{sec:simulation}
\subsection{Detection Rate of FRBs in the SKA era}
\label{subsec:FRB-rate}
Thanks to their extremely high event rate \citep{2013Sci...341...53T}, abundant FRBs are expected 
to be detected by the upcoming telescopes. Note that uncertainties in the fluence distribution and spectral 
index of the FRB population would make the estimate of the event detection rate for a specific telescope 
uncertain by several orders of magnitude \citep{2016ApJ...830...75V,2019MNRAS.483.1342J,2023SCPMA..6620412Z}. 
However, it is feasible to give a simple calculation on the event detection rate at operating frequencies 
close to those at which FRB detections are conducted (i.e., 1.2--1.7 GHz, as probed by the SKA's mid-frequency 
instrument), based on reasonable assumptions on the slope of the differential fluence distribution (i.e., 
the $\log N-\log F$ distribution). The cumulative all-sky event rate $N_{\rm sky}$ of FRBs above a given 
fluence threshold $F_{\nu}$ is generally described as a power-law model \citep{2018MNRAS.480.4211M},
\begin{equation}
N_{\rm sky}(>F_{\nu})=N'_{\rm sky}\left(\frac{F_{\nu}}{F'_{\nu}}\right)^{\alpha}\;,
\label{eq:Nsky}
\end{equation}
where $N'_{\rm sky}$ is the known event rate above the fluence threshold $F'_{\nu}$ estimated by existing 
FRB surveys, and the power-law index $\alpha=-1.5$ is in line with the expectation of a non-evolving
population in Euclidean space \citep{2016ApJ...830...75V}.

The mid-frequency aperture array of the first phase of SKA (SKA1-MID) would reach a $10\sigma$ sensitivity 
of $\mathrm{14\,mJy}$ from 0.9 to 1.67 $\mathrm{GHz}$ for an integration time of 1 ms \citep{2015aska.confE..51F}. 
While the actual scope of the second phase of SKA (SKA2) will eventually be subject to further discussions, 
the performance of SKA2 may be 10 times the SKA1 sensitivity in the frequency range of 0.35--24
$\mathrm{GHz}$.\footnote{\url{https://www.skao.int/en/science-users/118/ska-telescope-specifications}} 
That is, the adopted  $10\sigma$ fluence threshold of SKA2-MID is $F_{\nu}=1.4$ $\mathrm{mJy\,ms}$. 
With Equation~(\ref{eq:Nsky}), we can estimate the all-sky event rate observed by SKA2-MID based on 
the rate measurements of Parkes and ASKAP telescopes. \cite{2018MNRAS.475.1427B} obtained
$N'_{\rm sky}=(1.7^{+1.5}_{-0.9})\times10^{3}\,\mathrm{sky^{-1}\,day^{-1}}$ above the fluence threshold of
$F'_{\nu}=2$ $\mathrm{Jy\,ms}$ from Parkes FRB surveys. \cite{2018Natur.562..386S} obtained
$N'_{\rm sky}=(37\pm8)\,\mathrm{sky^{-1}\,day^{-1}}$ above the fluence threshold of $F'_{\nu}=26$ $\mathrm{Jy\,ms}$ 
from ASKAP FRB surveys. Based on the Parkes result, the estimated all-sky event rate for SKA2-MID is
\begin{equation}
N_{\rm sky}=\left(9.2^{+8.1}_{-4.9}\right)\times10^{7}\,\mathrm{sky^{-1}\,day^{-1}}\;,
\end{equation}
and according to the ASKAP result, we have
\begin{equation}
N_{\rm sky}=\left(9.4\pm2.0\right)\times10^{7}\,\mathrm{sky^{-1}\,day^{-1}}\;.
\end{equation}
It is obvious that the two estimates are in good agreement with each other.

Considering the solid angle $\Delta \Omega$ covered on the sky by the SKA instantaneous observation, 
the expected number of FRBs can be estimated by
\begin{equation}
N_{\rm exp}=\frac{\Delta \Omega}{4 \pi}T N_{\rm sky}\;,
\end{equation}
where $T$ is the exposure time on source. In general, the overall sky coverage of SKA2-MID would reach
$\sim25,000-30,000\,\mathrm{deg^{2}}$ \citep{2016arXiv161000683T}. Since FRBs are transient phenomena, 
here we conservatively take $\Delta \Omega\approx20\,\mathrm{deg^{2}}$ $(\sim 0.006\,\mathrm{sr})$ 
as the effective instantaneous field of view of SKA2-MID \citep{2010PASA...27..272M}. Following 
\cite{2023SCPMA..6620412Z}, we assume that SKA2-MID would spend an average of 20\% of observing time 
per year on FRB search, and take $T=20\% \times 365\,\mathrm{day\,yr^{-1}}$. On the basis of the Parkes 
and ASKAP results, the expected detection rates of FRBs by SKA2-MID are then
$\sim(3.3^{+2.9}_{-1.7})\times10^{6}\,\mathrm{events\,yr^{-1}}$
and $\sim (3.3\pm0.7)\times10^{6}\,\mathrm{events\,yr^{-1}}$, respectively. Thus, we conclude that 
$\mathcal{O}(10^6)$ FRBs can be detected by SKA2-MID in a one-year observation.

The precise localization of FRBs is crucial for FRB cosmology. To obtain their host galaxy redshifts 
via optical follow-up, the positions of FRBs should be localized to within one arcsecond \citep{2015aska.confE..55M}.
When complete, the SKA will have the ability of localizing FRBs to an accuracy of sub-arcseconds, 
and thus it is reasonable to expect that all the host galaxies of the SKA-detected FRBs would be 
identified and their redshifts would be measured.

\subsection{Mock FRB Data}
\label{subsec:mock}
In this work, we perform Monte Carlo simulations to explore the prospect of \he reionization detection 
with future FRB measurements in the SKA era. The fiducial flat $\Lambda$CDM model with cosmological 
parameters derived from Planck observations ($H_{0}=67.4$ km $\rm s^{-1}$ $\rm Mpc^{-1}$, $\Omega_{\rm m}=0.315$,
and $\Omega_{b}=0.0493$; \citealt{2020AA...641A...6P}) is adopted to simulate FRB data. 
The fiducial values of the reionization parameters are taken to be $A_{\rm He}=1$ and $z_{\rm re}=3$,
though we will discuss the effect of different fiducial $z_{\rm re}$ in the following subsection.

The redshift distribution of FRBs is expressed as
\begin{equation}
P(z)\propto \frac{4 \pi D_{C}^{2}(z)cR(z)}{H(z)\left(1+z\right)}\;,
\label{eq:z-distribution}
\end{equation}
where $D_{C}(z)=c\int_{0}^{z}1/H(z)\mathrm{d}z$ is the comoving distance, $H(z)$ is the Hubble parameter, 
and $R(z)$ denotes the intrinsic event rate density distribution. So far, the actual $R(z)$ distribution 
of FRBs is poorly known. \cite{2021PhRvD.103h3536Q} stressed that the assumed redshift distributions 
(i.e., the assumed $R(z)$ distributions) have significant effect on cosmological parameter estimation.
The true redshift distribution of FRBs is inextricably linked to the corresponding progenitor system(s) 
of FRBs, and it is commonly assumed that the FRB rate follows the cosmic star formation history (SFH) 
or compact star mergers. In the merger model, a compact star binary system needs to experience a long 
inspiral phase before the final merger, so there is a significant time delay with respect to star 
formation. In the literature, three types of merger delay timescale models have been discussed
\citep{2011ApJ...727..109V,2015ApJ...812...33S,2015MNRAS.448.3026W}: Gaussian delay model, lognormal 
delay model, and power-law delay model. Thus, we investigate four FRB event rate density models in detail:

\begin{itemize}
    \item SFH model: we adopt the approximate analytical form derived by \cite{2008ApJ...683L...5Y} 
using a wide range of the star formation rate data
\begin{equation}
R(z)=\left[ \left(1+z\right)^{3.4\eta} + \left(\frac{1+z}{5000}\right)^{-0.3\eta} + \left(\frac{1+z}{9}\right)^{-3.5\eta} \right]^{1/\eta}\;,
\end{equation}
where $\eta=-10$.
\end{itemize}

\begin{itemize}
    \item Compact star merger models: for the Gaussian delay model, we adopt the empirical formula 
    \citep{2011ApJ...727..109V}
\begin{align}
 R(z)= & \left[ \left(1+z\right)^{5.0\eta} + \left(\frac{1+z}{0.17}\right)^{0.87\eta}\right. \nonumber\\ 
& \left. + \left(\frac{1+z}{4.12}\right)^{-8.0\eta} + \left(\frac{1+z}{4.05}\right)^{-20.5\eta} \right]^{1/\eta}\;,
\end{align}
where $\eta=-2$.
\end{itemize}

\begin{itemize}
   \item Compact star merger models: for the lognormal delay model \citep{2015MNRAS.448.3026W}, 
   the empirical formula of $R(z)$ is 
\begin{align}
 R(z)= & \left[ \left(1+z\right)^{5.7\eta} + \left(\frac{1+z}{0.36}\right)^{1.3\eta}\right. \nonumber\\ 
& \left. + \left(\frac{1+z}{3.3}\right)^{-9.5\eta} + \left(\frac{1+z}{3.3}\right)^{-24.5\eta} \right]^{1/\eta}\;,
\end{align}
where $\eta=-2$.
\end{itemize}

\begin{itemize}
   \item Compact star merger models: for the power-law delay model \citep{2015MNRAS.448.3026W}, one has
\begin{align}
 R(z)= & \left[ \left(1+z\right)^{1.9\eta} + \left(\frac{1+z}{2.5}\right)^{-1.2\eta}\right. \nonumber\\ 
& \left. + \left(\frac{1+z}{3.8}\right)^{-4.4\eta} + \left(\frac{1+z}{7.7}\right)^{-11\eta} \right]^{1/\eta}\;,
\end{align}
where $\eta=-2.6$.
\end{itemize}

\begin{figure}
\begin{center}
\vskip-0.1in
\includegraphics[width=0.47\textwidth]{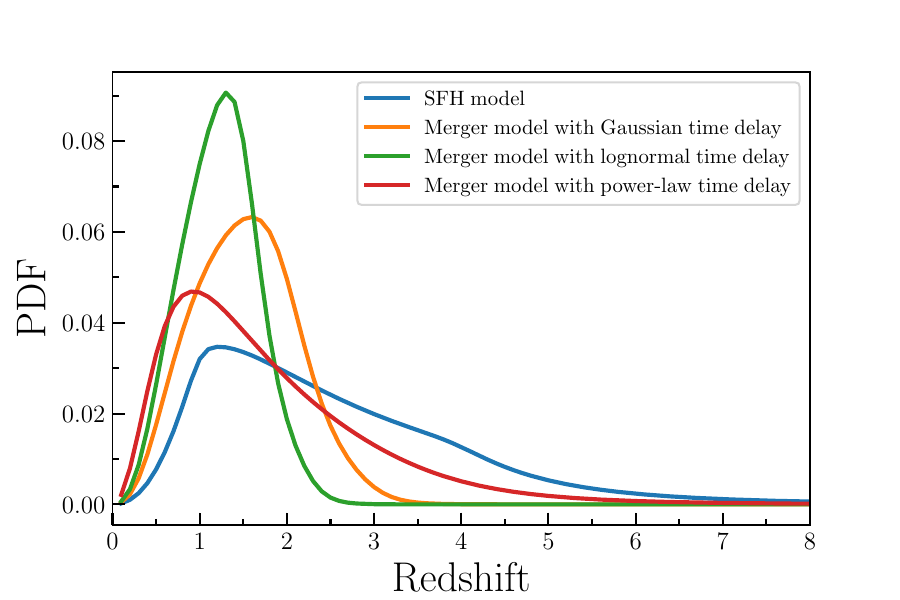}
\caption{PDFs of redshifts derived from the SFH model and three compact star merger models with different 
merger delay time distributions (Gaussian, lognormal, and power-law).}
\label{fig2}
\vskip-0.2in
\end{center}
\end{figure}

Figure~\ref{fig2} presents the FRB redshift distributions expected from all four assumed models. It is 
clear that the SFH model has the widest $z$ distribution. Due to the merger delay, the other three merger 
models have narrower $z$ distributions, with the overall redshift gradually shifting to lower redshifts 
in the order of power-law, Gaussian, and lognormal models \citep{2021MNRAS.501..157Z}. Note that \he 
reionization is generally expected to occur at $z\sim3$ (e.g., \citealt{2011MNRAS.410.1096B}). 
In order to effectively detect the signature of \he reionization, many high-$z$ FRBs on either side of 
the expected reionization redshift are required. However, only low-$z$ ($z\leq 3$) FRBs are predicted by 
the Gaussian and lognormal delay models (see Figure~\ref{fig2}). For the rest of this paper, we shall 
therefore only consider two cases: the FRB redshift distribution tracing the SFH and power-law delay models,
respectively.

\begin{figure}
\begin{center}
\vskip-0.2in
\includegraphics[width=0.45\textwidth]{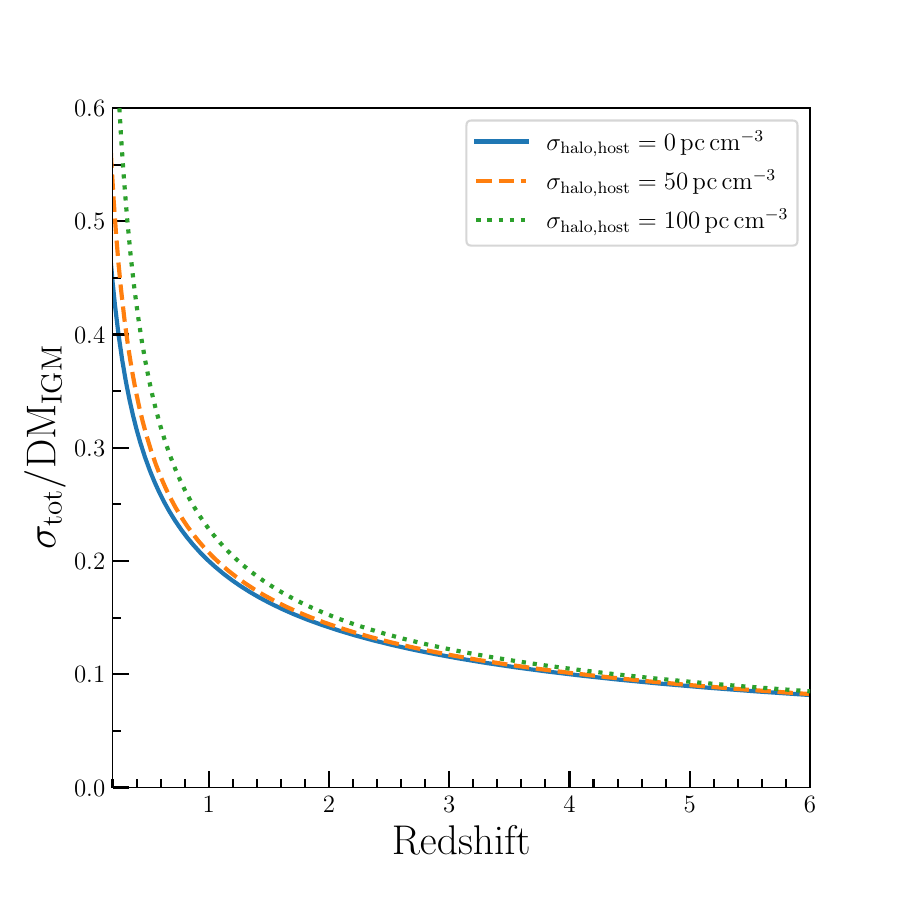}
\vskip-0.1in
\caption{The fractional uncertainty on a measured $\mathrm{DM_{IGM}}$, contributed by the IGM scatter 
($\sigma_{\mathrm{IGM}}$) and the uncertainty of the DM contributions from both the Milky Way halo 
and the FRB host galaxy ($\sigma_{\mathrm{halo,\,host}}$), as a function of redshift. Curves show 
the quadrature sums of the two contributions for three cases with different $\sigma_{\mathrm{halo,\,host}}$
values: $0$ $\mathrm{pc\,cm^{-3}}$ (blue solid line), $50$ $\mathrm{pc\,cm^{-3}}$ (yellow dashed line), 
and $100$ $\mathrm{pc\,cm^{-3}}$ (green dotted line).}
\label{fig3}
\vskip-0.2in
\end{center}
\end{figure}

In our simulations, the true FRB host galaxy redshifts $z^{\rm true}$ are randomly generated 
from the probability density functions (PDFs) of Equation~(\ref{eq:z-distribution}) for the SFH and 
power-law delay cases. After generating $z^{\rm true}$, we simulate the measured redshifts $z^{\rm mea}$
by assuming an uncertainty of 10\% on the spectroscopic data of host galaxies (as expected from
the James Webb Space Telescope). That is, we draw $z^{\rm mea}$ from a Gaussian distribution 
$\mathcal{N}(\mu_{z}=z^{\rm true},\;\sigma_{z}=0.1z^{\rm true})$ \citep{2022ApJ...933...57H,2024arXiv240805722M}.

With the true source redshift $z^{\rm true}$, 
we infer the fiducial value of $\mathrm{DM^{fid}_{IGM}}$ from Equation~(\ref{eq:SNR}). Following
previous works \citep{2019PhRvD.100h3533K,2020PhRvD.101j3019L}, the line-of-sight fluctuation of 
$\mathrm{DM_{IGM}}$ ($\sigma_{\mathrm{IGM}}$) is randomly added to the fiducial value of 
$\mathrm{DM^{fid}_{IGM}}$, assuming
\begin{equation}
\sigma_{\mathrm{IGM}}=20\% \frac{\mathrm{DM^{fid}_{IGM}}}{\sqrt{z^{\rm true}}}\;.
\end{equation}
That is, we sample the $\mathrm{DM^{mea}_{IGM}}$ measurement according to the Gaussian distribution 
${\rm DM_{IGM}^{mea}}=\mathcal{N}({\rm DM_{IGM}^{fid}},\;\sigma_{\rm IGM})$. We note that
the observed $\mathrm{DM_{obs}}$ has contributions from the interstellar medium ($\mathrm{DM_{MW}}$)
and dark matter halo ($\mathrm{DM_{halo}}$) in the Milky Way, the IGM ($\mathrm{DM_{IGM}}$), and 
the FRB host galaxy ($\mathrm{DM_{host}}$). The $\mathrm{DM_{IGM}}$ value can then be derived 
by deducting $\mathrm{DM_{MW}}$, $\mathrm{DM_{halo}}$, and $\mathrm{DM_{host}}$ to $\mathrm{DM_{obs}}$.
While the corresponding uncertainty of $\mathrm{DM_{IGM}}$ should include the scatter due to 
the inhomogeneous IGM and the scatter caused by distinctive host properties and halo, 
the latter scatter (order of $\mathrm{10\,pc\,cm^{-3}}$) is much smaller than the former one 
(order of $\mathrm{10^{2}\,pc\,cm^{-3}}$) at the reionization epoch (e.g., \citealt{2019MNRAS.485..648P}).
Figure~\ref{fig3} shows the evolution of the fractional uncertainty on a measured $\mathrm{DM_{IGM}}$, 
accounting for the line-of-sight fluctuation of $\mathrm{DM_{IGM}}$ ($\sigma_{\mathrm{IGM}}$) 
and the uncertainty of both $\mathrm{DM_{halo}}$ and $\mathrm{DM_{host}}$ ($\sigma_{\mathrm{halo,\,host}}$), 
i.e., $\sigma_{\mathrm{tot}}=(\sigma_{\mathrm{IGM}}^{2}+\sigma^{2}_{\mathrm{halo,\,host}})^{1/2}$.
At low redshift, $\mathrm{DM_{IGM}}$ is small, so the fractional uncertainty $\sigma_{\mathrm{tot}}/\mathrm{DM_{IGM}}$ 
is relatively large. Since lines of sight average over IGM inhomogeneity and $\mathrm{DM_{IGM}}$ increases
at higher redshift, $\sigma_{\mathrm{tot}}/\mathrm{DM_{IGM}}$ gradually decreases. Moreover, 
one can see from Figure~\ref{fig3} that adding the halo and host uncertainty (with the value of 
$\sigma_{\mathrm{halo,\,host}}=0,\,50,$ or $100$ $\mathrm{pc\,cm^{-3}}$) in quadrature gives
negligible effect on $\sigma_{\mathrm{tot}}/\mathrm{DM_{IGM}}$, especially around the expected
reionization redshift $z_{\rm re}\ga 3$. Therefore, we neglect the uncertainty of both 
$\mathrm{DM_{halo}}$ and $\mathrm{DM_{host}}$ subtractions from $\mathrm{DM_{obs}}$ to derive
$\mathrm{DM_{IGM}}$ observationally. 

Using the method described above, one can generate a catalog of the simulated FRB events with 
$z^{\rm mea}$, $\mathrm{DM^{mea}_{IGM}}$, and $\sigma_{\mathrm{IGM}}$. Due to the lack of
computational efficiency, we first simulate a population of $10^{4}$ such events, assuming that 
the FRB redshift distribution follows the SFH model or the power-law delay model separately.

\subsection{Results}
For a set of $10^{4}$ simulated FRBs, we only employ those FRBs with $z<6$ for \he reionization
parameter estimation, since the ionization fractions of hydrogen $X_{e,\mathrm{H\,II}}(z)$ at 
$z\ge6$ are not well understood and we take all of the hydrogen to be ionized at $z=6$. 
We then utilize the standard Bayes theorem to explore the posterior probability distribution
$\mathcal{P}(\mathbf{\theta}|\mathcal{D})$ of the model parameters $\mathbf{\theta}$, provided 
the observed dataset $\mathcal{D}$, i.e.,
\begin{equation}
\mathcal{P}(\mathbf{\theta}|\mathcal{D})=\frac{\mathcal{L}(\mathcal{D}|\mathbf{\theta})\,\pi(\mathbf{\theta})}{\mathcal{P}(\mathcal{D})}\;,
\end{equation}
where $\mathcal{L}(\mathcal{D}|\mathbf{\theta})$ is the likelihood of data conditional on the hypothetical model, 
$\pi(\mathbf{\theta})$ denotes some prior knowledge about the model parameters, and $\mathcal{P}(\mathcal{D})$ 
is the Bayesian evidence (which can be regarded as the normalized parameter and has no effect on our analysis). 
In our study, $\mathcal{D}$ is the mock dataset consisting of $\mathrm{DM^{mea}_{IGM}}$ and $z^{\rm mea}$, 
and the free parameters to be constrained are $\mathbf{\theta}=\{A_{\rm He},\,z_{\rm re}\}$. The likelihood 
is given by \citep{2024arXiv240805722M}
\begin{align}
\mathcal{L}(\mathbf{\theta})= & \exp \left[-\frac{1}{2}\sum_{i}\left(\frac{\mathrm{DM^{mea}_{IGM,\emph{i}}}-\mathrm{DM^{th}_{IGM}}(z^{\rm true}_{i};\,\mathbf{\theta})}{\sigma_{\mathrm{IGM},i}}\right)^{2}\right] \nonumber\\ 
& \times \exp \left[-\frac{1}{2}\sum_{i}\left(\frac{z^{\rm mea}_{i}-z^{\rm true}_{i}}{\sigma_{z,i}}\right)^{2}\right] \;,
\end{align}
where $\mathrm{DM^{th}_{IGM}}(z_{i};\,\mathbf{\theta})$ is the theoretical value of $\mathrm{DM_{IGM}}$ 
calculated from the set of \he reionization parameters $\mathbf{\theta}$ (see Equation~\ref{eq:SNR}).
To ensure the final resulting constraints are unbiased, the simulation process is repeated 100 times 
for each FRB data set by using different noise seeds. 

\begin{figure}
\vskip-0.2in
\centerline{\includegraphics[width=0.55\textwidth]{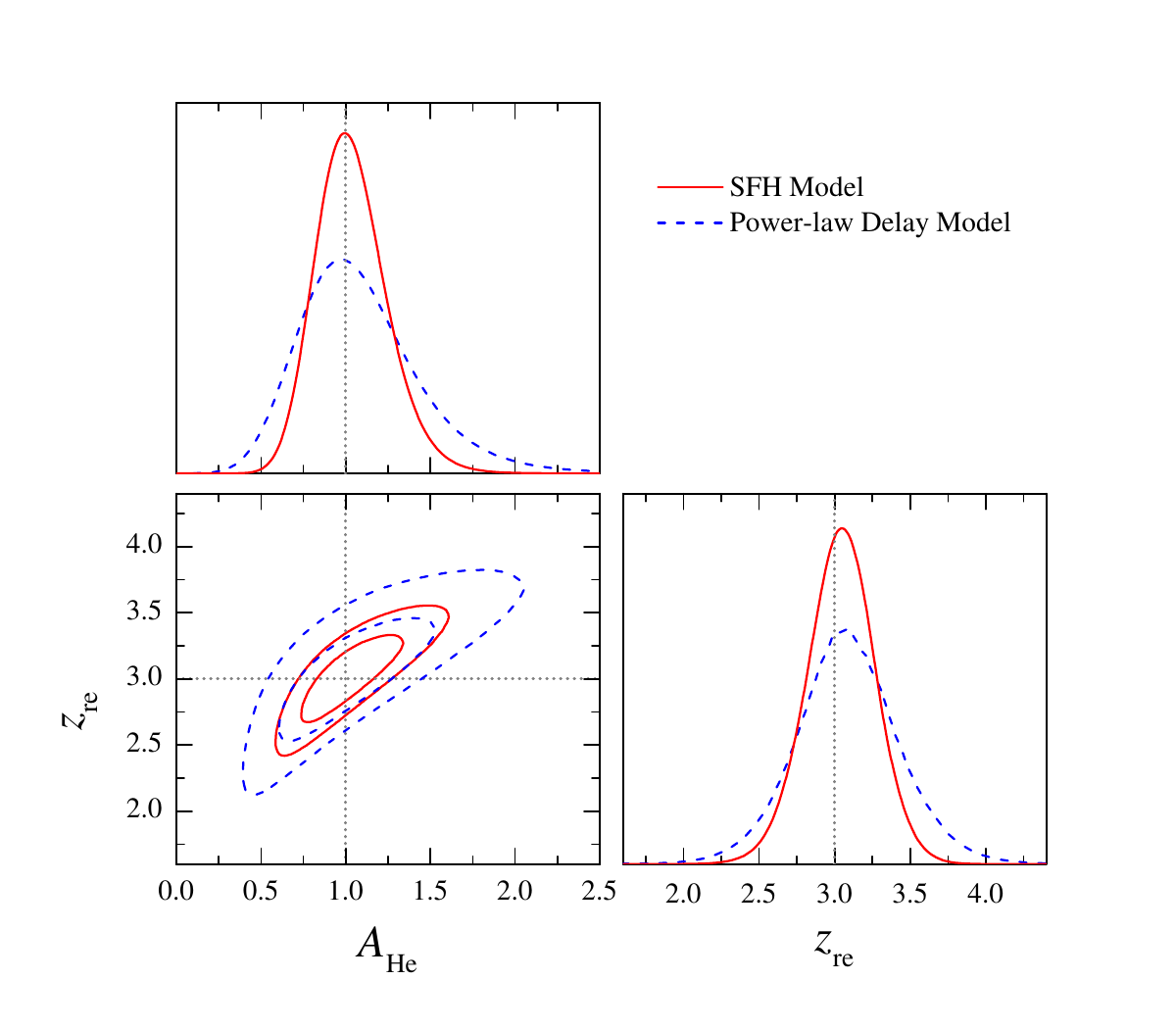}}
\vskip-0.1in
\caption{1-D marginalized probability distributions and $1-2\sigma$ constraint contours for
the reionization parameters $A_{\rm He}$ and $z_{\rm re}$, using $10^{4}$ simulated FRB events. 
The solid and dashed contours correspond to the cases of $z$ distribution tracing the SFH model 
and the compact star merger model with power-law merger delay time distribution, respectively.
The vertical dotted lines mark the input parameter values used to generate the FRB mocks.}
\label{fig4}
\end{figure}

Figure~\ref{fig4} displays the resulting constraints on the reionization parameters $A_{\rm He}$ 
and $z_{\rm re}$ from $10^{4}$ simulated FRBs. For the case of $z$ distribution tracing the SFH 
model (solid contours), we find the marginalized parameter uncertainties are $\sigma(A_{\rm He})=0.20$ 
and $\sigma(z_{\rm re})=0.22$. That is, the detection $\mathrm{S/N}$ for \he reionization is 
$\mathrm{S/N}=A_{\rm He}/\sigma(A_{\rm He})=1.00/0.20=5.0$. Besides, the sudden reionization 
redshift can be inferred to be $z_{\rm re}=3.00\pm0.22$. To investigate the effect of $z$ 
distribution, in Figure~\ref{fig4} we also plot those constraints obtained from the case of 
$z$ distribution tracing the power-law delay model (dashed contours). Now we have  
$A_{\rm He}=0.99\pm0.31$ and $z_{\rm re}=2.99\pm0.31$. The detection $\mathrm{S/N}$ decreases to
$\mathrm{S/N}=A_{\rm He}/\sigma(A_{\rm He})=0.99/0.31=3.2$. We can see that the constraint 
precisions of $A_{\rm He}$ and $z_{\rm re}$ obtained from the power-law delay model are 
slightly worse than those from the SFH model. This is mainly because reionization parameters 
are more sensitive to high-$z$ FRB data, and more high-$z$ mock events, especially around 
the expected reionization redshift $z_{\rm re}\ga 3$, are generated in the SFH model 
(see Figure~\ref{fig2}).

\begin{table*}
\tabcolsep=0.7cm
\centering \caption{Summary of Reionization Parameter Constraints from
$N$ Simulated FRB Events Using Different $z$ Distributions}
\begin{tabular}{lcccc}
\hline
\hline
 & \multicolumn{2}{c}{SFH Model}  & \multicolumn{2}{c}{Power-law Delay Model} \\
 \cline{2-5} 
 $N\,(\times10^{4})$ &  $A_{\rm He}$  &  $z_{\rm re}$ &  $A_{\rm He}$  &  $z_{\rm re}$\\
\hline
1   &   $1.00\pm0.20$   &   $3.00\pm0.22$   &   $0.99\pm0.31$   &   $2.99\pm0.31$  \\
2   &   $1.01\pm0.14$   &   $3.02\pm0.15$   &   $1.01\pm0.21$   &   $3.00\pm0.21$\\
3   &   $1.00\pm0.11$   &   $3.00\pm0.12$   &   $1.00\pm0.17$   &   $3.00\pm0.17$\\
4   &   $1.00\pm0.10$   &   $3.00\pm0.11$   &   $1.00\pm0.15$   &   $3.00\pm0.15$\\
5   &   $1.00\pm0.09$   &   $3.00\pm0.10$   &   $1.01\pm0.13$   &   $3.00\pm0.13$\\
\hline
\end{tabular}
\label{table1}
\end{table*}

\begin{figure}
\centerline{\includegraphics[width=0.55\textwidth]{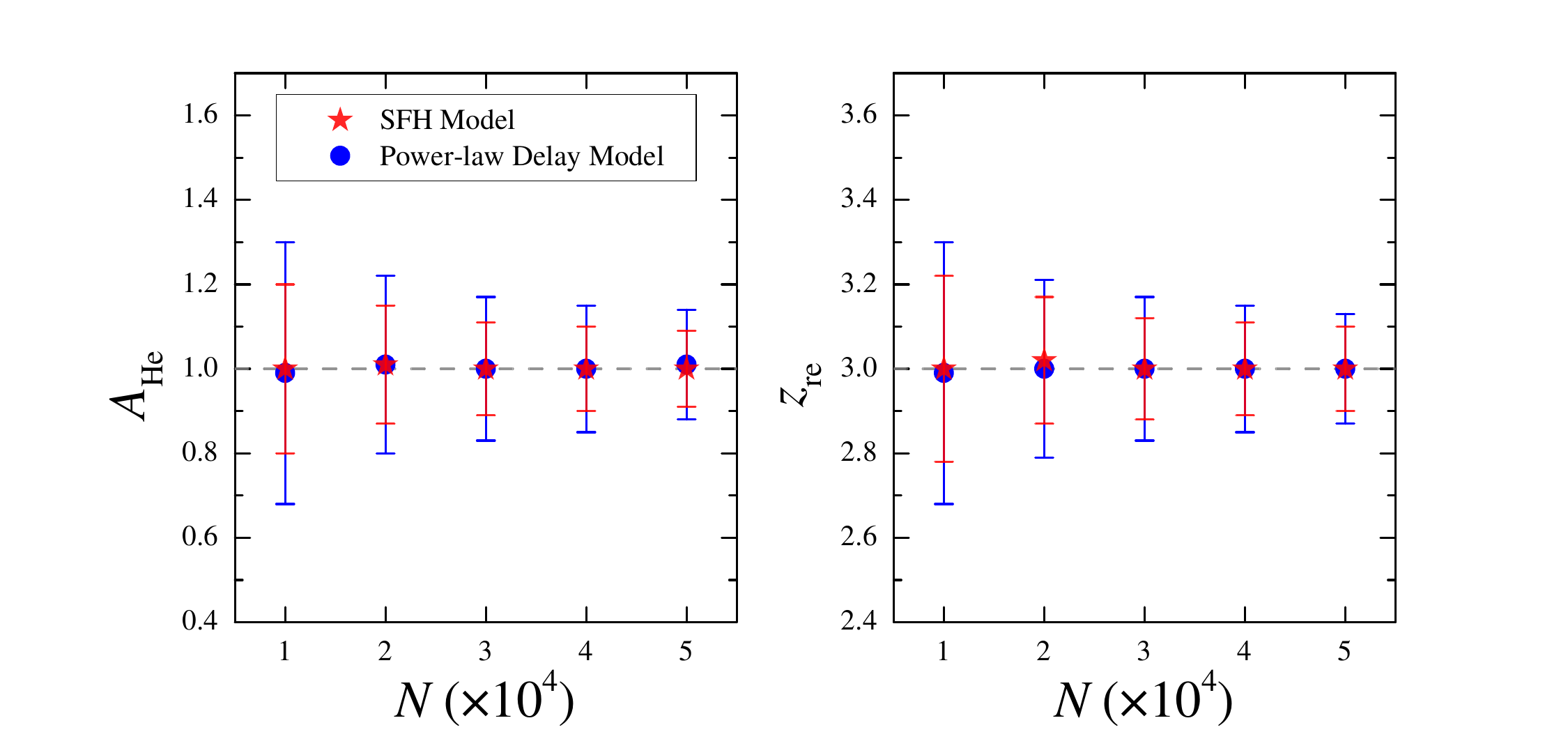}}
\caption{Best-fit reionization parameters ($A_{\rm He}$ and $z_{\rm re}$) and their corresponding 
$1\sigma$ uncertainties as a function of the number of simulated FRB data. The star
and circle symbols correspond to the cases of $z$ distribution tracing the SFH and power-law delay
models, respectively. The horizontal dashed lines stand for the fiducial values.}
\label{fig5}
\end{figure}

In order to comprehensively test what role the number of simulated FRBs ($N$) could play in 
\he reionization detection, we show the best-fit reionization parameters and their corresponding 
$1\sigma$ uncertainties as a function of $N$ in Figure~\ref{fig5} and Table~\ref{table1}. 
We find that the uncertainties of these constraints from the FRB data are well 
consistent with the $1/\sqrt{N}$ behavior, independent of what kind of $z$ distribution is assumed. 
According to our estimation of the FRB detection rate by SKA2-MID in Section~\ref{subsec:FRB-rate},
the one-year SKA2 observation would detect $\mathcal{O}(10^6)$ FRBs. In principle, it would be
ideal to show the analysis with $10^6$ mock FRBs. However, due to the lack of computational efficiency, 
we estimate the capacity of $10^6$ mocks by extrapolating from present results of $10^{4}$ mocks. 
Since all uncertainties will scale in proportion to the $1/\sqrt{N}$ behavior, $10^6$ FRB data 
can give the reionization parameter uncertainties $\sigma(A_{\rm He})=0.020$ and 
$\sigma(z_{\rm re})=0.022$ ($\sigma(A_{\rm He})=0.031$ and $\sigma(z_{\rm re})=0.031$)
for the case of $z$ distribution tracing the SFH (power-law delay) model. Hence, for a one-year 
SKA2 observation with $10^6$ FRBs, the detection $\mathrm{S/N}$ can approach 32--50.

\begin{figure*}
\begin{center}
\vskip-0.2in
\includegraphics[width=0.45\textwidth]{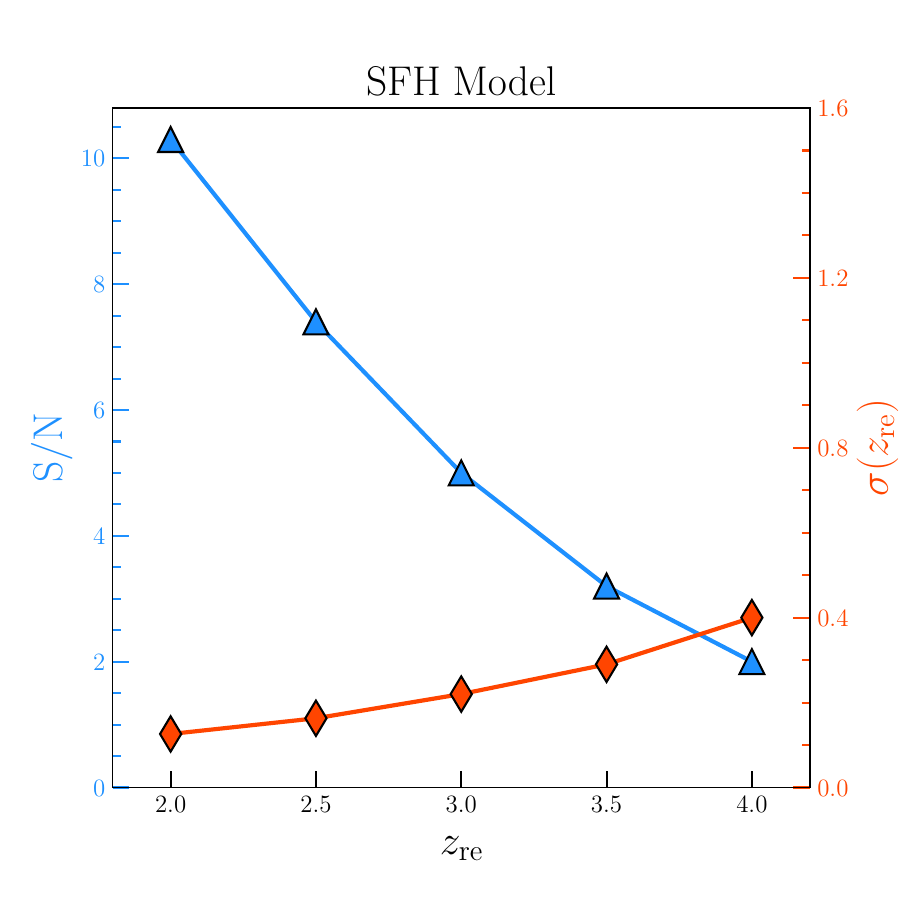}
\includegraphics[width=0.45\textwidth]{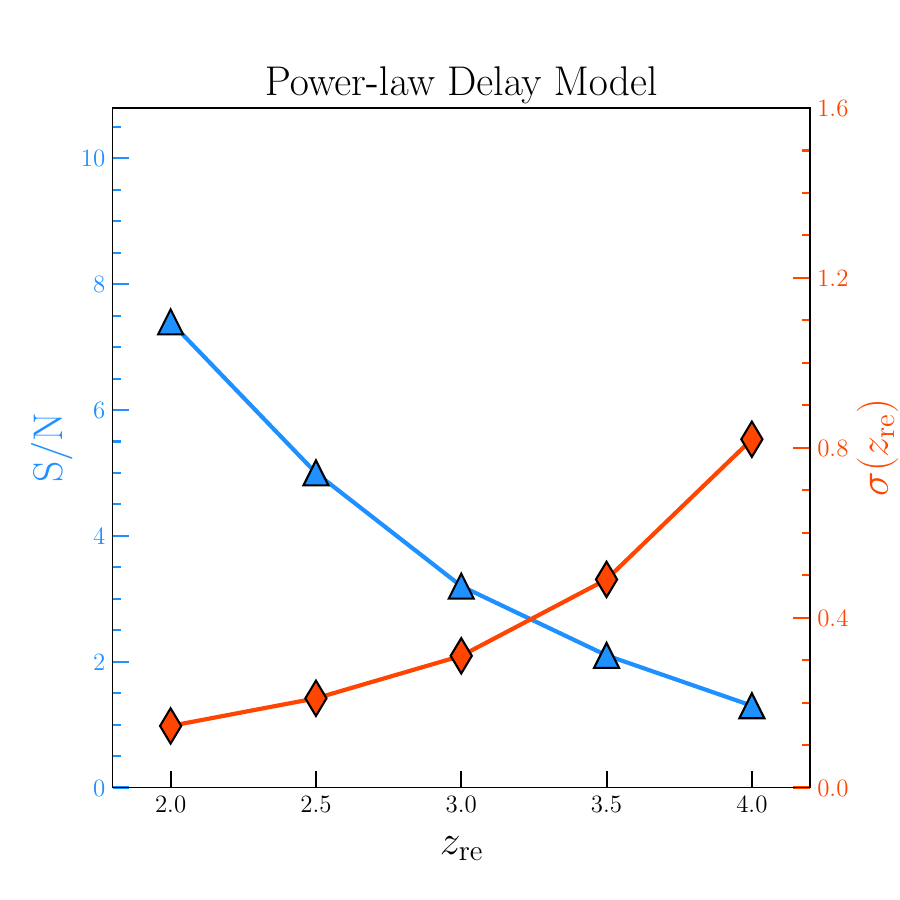}
\vskip-0.1in
\caption{Sensitivity of the signal-to-noise ratio $\mathrm{S/N}=A_{\rm He}/\sigma(A_{\rm He})$ 
for the detection of \he reionization and the uncertainty on the reionization redshift 
$\sigma(z_{\rm re})$ to the fiducial value of $z_{\rm re}$. All the constraints are derived from 
$10^{4}$ simulated FRBs for the cases of $z$ distribution following the SFH model (left panel) 
and the power-law delay model (right panel).}
\label{fig6}
\vskip-0.2in
\end{center}
\end{figure*}

In our above simulations, the fiducial reionization redshift is set to be $z_{\rm re}=3$.
To study the effect of different fiducial $z_{\rm re}$, we estimate the sensitivity as 
we vary $z_{\rm re}$. Note that here we also mock a population of $10^{4}$ FRBs with
the $z$ distribution following the SFH and power-law delay models, respectively. 
Figure~\ref{fig6} shows that no matter which $z$ distribution is assumed, the $\mathrm{S/N}$ 
gets better rapidly with lower $z_{\rm re}$, but falls below $\mathrm{S/N}=1$ for $z_{\rm re}\ga 4$.
As the fiducial value of $z_{\rm re}$ grows, the reionization redshift uncertainty 
$\sigma(z_{\rm re})$ becomes larger. These are pretty easy to understand, because
the number of FRBs at higher $z_{\rm re}$ is smaller (see the blue and red lines in Figure~\ref{fig2}).
Over the main range of the expected reionization redshifts, say $z_{\rm re}\approx 2.5-3.5$,
the $\mathrm{S/N}$ varies from 7.4 (5.0) to 3.2 (2.1), and $\sigma(z_{\rm re})$ varies from 
0.16 (0.21) to 0.29 (0.49) for the SFH (power-law delay) model.

In our analysis, we have neglected the uncertainty of the DM contributions from 
both the Milky Way halo and the FRB host ($\sigma_{\mathrm{halo,\,host}}$), as described in 
Section~\ref{subsec:mock}. Although it seems to be a reasonable treatment as suggested by 
Figure~\ref{fig3}, it would be good to add an analysis incorporating $\sigma_{\mathrm{halo,\,host}}$
and check if there is any bias in the posterior recovery. Especially, when \he reionization
is assumed to occur at $z=2$, adding the halo and host uncertainty (for the maximum case of
$\sigma_{\mathrm{halo,\,host}}=100$ $\mathrm{pc\,cm^{-3}}$) in quadrature gives a slight increasement
of $\sigma_{\mathrm{tot}}/\mathrm{DM_{IGM}}$ (see Figure~\ref{fig3}). To investigate how sensitive 
our results on the uncertainties of $A_{\rm He}$ and $z_{\rm re}$ are on the choice of $\sigma_{\mathrm{halo,\,host}}$, 
we also perform two parallel comparative analyses of $10^{4}$ simulated FRBs using $\sigma_{\mathrm{halo,\,host}}=0$ 
$\mathrm{pc\,cm^{-3}}$ and $100$ $\mathrm{pc\,cm^{-3}}$, assuming that the fiducial reionization redshift 
is $z_{\rm re}=2$ and the $z$ distribution traces the power-law delay model. If we change 
$\sigma_{\mathrm{halo,\,host}}$ from $0$ $\mathrm{pc\,cm^{-3}}$ to $100$ $\mathrm{pc\,cm^{-3}}$,
then $\sigma(A_{\rm He})$ goes from $0.135$ to $0.144$, and $\sigma(z_{\rm re})$ from $0.145$ to $0.154$.
So there is no concern regarding a possibly large dependence on $\sigma_{\mathrm{halo,\,host}}$.

We emphasize that all our results are based on the assumption of sudden \he reionization.
However, in reality, it is expected that the reionization phenomenon should evolve gradually with $z$. 
Assuming a linear evolution in redshift for the \he reionization within a range of $z_{\rm re,min}$ 
to $z_{\rm re,max}$, \cite{2021PhRvD.103j3526B} found that the difference between results for 
sudden reionization and gradual reionization is generally not significant. This is a positive 
outcome in the sense that our results on \he reionization detection are robust to the assumption 
about suddenness, but does imply that the duration of the reionization process is difficult to determine.

\section{Summary and discussion}
\label{sec:sum}
Cosmic reionization typically refers to the ionization of neutral hydrogen and helium at
redshifts $z\ga6$. The singly ionized helium, \mbox{He {\sc ii}}, is expected to have 
undergone the same phase transition, but at a relatively lower redshift, and is often 
referred to as \he reionization. Compared to the former one related to neutral hydrogen 
and helium reionization above $z\sim6$, the later \he reionization is more accessible to 
observations by next generation surveys. FRBs, with their short pulse nature, high event rate, 
and large cosmological distance, are a promising tool for probing the epoch of \he reionization.

In this work, we have investigated the prospect of \he reionization detection with future 
FRB data to be detected by the upcoming SKA. With the definition of a simple $\mathrm{S/N}$ 
criterion for detecting \he reionization, we performed Monte Carlo simulations to assess 
the ability of the SKA-era FRB observation to constrain the detection $\mathrm{S/N}$ and 
the redshift $z_{\rm re}$ at which reionization occurs. In our simulations, we considered 
two different FRB redshift distributions: one is assumed to follow the SFH model, and 
the other one traces the compact star merger model 
with power-law merger delay time distribution. Based on a mock catalog of $10^{4}$ FRBs,
our results showed that the detection $\mathrm{S/N}\approx 3.2-5.0$ and the uncertainty of 
the reionization redshift $\sigma(z_{\rm re})\approx 0.22-0.31$, depending on the assumed 
$z$ distribution. Moreover, we confirmed that the uncertainties of the marginalized
reionization parameters are in good agreement with the $1/\sqrt{N}$ behavior, with $N$
the number of simulated FRBs. According to the existing FRB surveys and the possible 
detection performance of SKA2-MID, we expected that $\mathcal{O}(10^6)$ FRBs would be
detected by SKA2 in a one-year observation. Since all the reionization parameter uncertainties 
will scale as $1/\sqrt{N}$, $10^6$ FRB data can give $\mathrm{S/N}\approx 32-50$ and
$\sigma(z_{\rm re})\approx 0.022-0.031$. That is, for a one-year SKA2 observation with 
$10^6$ FRBs, the $\mathrm{S/N}$ for the detection of \he reionization can approach the 
$32-50\sigma$ level. We also inspected how the fiducial value of the reionization redshift
$z_{\rm re}$ affect the forecasts for \he reionization detection, and found this effect
was modest within the main range of the expected reionization redshifts.

It is worth highlighting that detecting high-$z$ FRBs is very important scientifically.
After the completion of SKA, an extensive dataset of high-$z$ FRBs might become available. 
This would provide a great opportunity to utilize the DM measurements of high-$z$ FRBs 
to directly probe the state of reionization in the IGM. Moreover, other world-leading 
radio survey telescopes such as the Hydrogen Intensity and Real-time Analysis eXperiment 
(HIRAX; \citealt{2016SPIE.9906E..5XN}), Deep Synoptic Array 2000-dishes (DAS-2000; 
\citealt{2019BAAS...51g.255H}), and Canadian Hydrogen Observatory and Radio-transient 
Detector (CHORD; \citealt{2019clrp.2020...28V}) are also anticipated to detect and 
localize tens of thousands of FRBs. With the rapid development of FRB detections, 
revealing the history of \he reionization with FRBs would be turned into reality.

\begin{acknowledgments}
We are grateful to the anonymous referee for constructive comments that 
have significantly improved our presentation.
This work is supported by the National SKA Program of China (2022SKA0130100), 
the National Natural Science Foundation of China (grant Nos. 12422307, 12373053, 12321003, and
12041306), the Key Research Program of Frontier Sciences (grant No. ZDBS-LY-7014)
of Chinese Academy of Sciences, International Partnership Program of Chinese Academy of Sciences
for Grand Challenges (114332KYSB20210018), the CAS Project for Young Scientists in Basic Research
(grant No. YSBR-063), and the Natural Science Foundation of Jiangsu Province (grant No. BK20221562).
\end{acknowledgments}

\bibliography{ms}{}

\begin{thebibliography}{}
\expandafter\ifx\csname natexlab\endcsname\relax\def\natexlab#1{#1}\fi
\providecommand{\url}[1]{\href{#1}{#1}}
\providecommand{\dodoi}[1]{doi:~\href{http://doi.org/#1}{\nolinkurl{#1}}}
\providecommand{\doeprint}[1]{\href{http://ascl.net/#1}{\nolinkurl{http://ascl.net/#1}}}
\providecommand{\doarXiv}[1]{\href{https://arxiv.org/abs/#1}{\nolinkurl{https://arxiv.org/abs/#1}}}

\bibitem[{{Barkana} \& {Loeb}(2001)}]{2001PhR...349..125B}
{Barkana}, R., \& {Loeb}, A. 2001, \physrep, 349, 125,
  \dodoi{10.1016/S0370-1573(01)00019-9}

\bibitem[{{Basu} {et~al.}(2024){Basu}, {Garaldi}, \&
  {Ciardi}}]{2024MNRAS.532..841B}
{Basu}, A., {Garaldi}, E., \& {Ciardi}, B. 2024, \mnras, 532, 841,
  \dodoi{10.1093/mnras/stae1488}

\bibitem[{{Becker} {et~al.}(2011){Becker}, {Bolton}, {Haehnelt}, \&
  {Sargent}}]{2011MNRAS.410.1096B}
{Becker}, G.~D., {Bolton}, J.~S., {Haehnelt}, M.~G., \& {Sargent}, W. L.~W.
  2011, \mnras, 410, 1096, \dodoi{10.1111/j.1365-2966.2010.17507.x}

\bibitem[{{Beniamini} {et~al.}(2021){Beniamini}, {Kumar}, {Ma}, \&
  {Quataert}}]{2021MNRAS.502.5134B}
{Beniamini}, P., {Kumar}, P., {Ma}, X., \& {Quataert}, E. 2021, \mnras, 502,
  5134, \dodoi{10.1093/mnras/stab309}

\bibitem[{{Bhandari} {et~al.}(2018){Bhandari}, {Keane}, {Barr}, {Jameson},
  {Petroff}, {Johnston}, {Bailes}, {Bhat}, {Burgay}, {Burke-Spolaor}, {Caleb},
  {Eatough}, {Flynn}, {Green}, {Jankowski}, {Kramer}, {Krishnan}, {Morello},
  {Possenti}, {Stappers}, {Tiburzi}, {van Straten}, {Andreoni}, {Butterley},
  {Chandra}, {Cooke}, {Corongiu}, {Coward}, {Dhillon}, {Dodson}, {Hardy},
  {Howell}, {Jaroenjittichai}, {Klotz}, {Littlefair}, {Marsh}, {Mickaliger},
  {Muxlow}, {Perrodin}, {Pritchard}, {Sawangwit}, {Terai}, {Tominaga}, {Torne},
  {Totani}, {Trois}, {Turpin}, {Niino}, {Wilson}, {Albert}, {Andr{\'e}},
  {Anghinolfi}, {Anton}, {Ardid}, {Aubert}, {Avgitas}, {Baret},
  {Barrios-Mart{\'\i}}, {Basa}, {Belhorma}, {Bertin}, {Biagi}, {Bormuth},
  {Bourret}, {Bouwhuis}, {Br{\^a}nza{\c{s}}}, {Bruijn}, {Brunner}, {Busto},
  {Capone}, {Caramete}, {Carr}, {Celli}, {Moursli}, {Chiarusi}, {Circella},
  {Coelho}, {Coleiro}, {Coniglione}, {Costantini}, {Coyle}, {Creusot},
  {D{\'\i}az}, {Deschamps}, {De Bonis}, {Distefano}, {Palma}, {Domi},
  {Donzaud}, {Dornic}, {Drouhin}, {Eberl}, {Bojaddaini}, {Khayati},
  {Els{\"a}sser}, {Enzenh{\"o}fer}, {Ettahiri}, {Fassi}, {Felis}, {Fusco},
  {Gay}, {Giordano}, {Glotin}, {Gregoire}, {Gracia-Ruiz}, {Graf}, {Hallmann},
  {van Haren}, {Heijboer}, {Hello}, {Hern{\'a}ndez-Rey}, {H{\"o}{\ss}l},
  {Hofest{\"a}dt}, {Hugon}, {Illuminati}, {James}, {de Jong}, {Jongen},
  {Kadler}, {Kalekin}, {Katz}, {Kie{\ss}ling}, {Kouchner}, {Kreter},
  {Kreykenbohm}, {Kulikovskiy}, {Lachaud}, {Lahmann}, {Lef{\`e}vre}, {Leonora},
  {Loucatos}, {Marcelin}, {Margiotta}, {Marinelli}, {Mart{\'\i}nez-Mora},
  {Mele}, {Melis}, {Michael}, {Migliozzi}, {Moussa}, {Navas}, {Nezri},
  {Organokov}, {P{\v{a}}v{\v{a}}la{\c{s}}}, {Pellegrino}, {Perrina},
  {Piattelli}, {Popa}, {Pradier}, {Quinn}, {Racca}, {Riccobene},
  {S{\'a}nchez-Losa}, {Salda{\~n}a}, {Salvadori}, {Samtleben}, {Sanguineti},
  {Sapienza}, {Sch{\"u}ssler}, {Sieger}, {Spurio}, {Stolarczyk}, {Taiuti},
  {Tayalati}, {Trovato}, {Turpin}, {T{\"o}nnis}, {Vallage}, {Van Elewyck},
  {Versari}, {Vivolo}, {Vizzocca}, {Wilms}, {Zornoza}, \&
  {Z{\'u}{\~n}iga}}]{2018MNRAS.475.1427B}
{Bhandari}, S., {Keane}, E.~F., {Barr}, E.~D., {et~al.} 2018, \mnras, 475,
  1427, \dodoi{10.1093/mnras/stx3074}

\bibitem[{{Bhattacharya} {et~al.}(2021){Bhattacharya}, {Kumar}, \&
  {Linder}}]{2021PhRvD.103j3526B}
{Bhattacharya}, M., {Kumar}, P., \& {Linder}, E.~V. 2021, \prd, 103, 103526,
  \dodoi{10.1103/PhysRevD.103.103526}

\bibitem[{{Caleb} {et~al.}(2019){Caleb}, {Flynn}, \&
  {Stappers}}]{2019MNRAS.485.2281C}
{Caleb}, M., {Flynn}, C., \& {Stappers}, B.~W. 2019, \mnras, 485, 2281,
  \dodoi{10.1093/mnras/stz571}

\bibitem[{{Cordes} \& {Chatterjee}(2019)}]{2019ARA&A..57..417C}
{Cordes}, J.~M., \& {Chatterjee}, S. 2019, \araa, 57, 417,
  \dodoi{10.1146/annurev-astro-091918-104501}

\bibitem[{{Dai} \& {Xia}(2021)}]{2021JCAP...05..050D}
{Dai}, J.-P., \& {Xia}, J.-Q. 2021, \jcap, 2021, 050,
  \dodoi{10.1088/1475-7516/2021/05/050}

\bibitem[{{Deng} \& {Zhang}(2014)}]{2014ApJ...783L..35D}
{Deng}, W., \& {Zhang}, B. 2014, \apj, 783, L35,
  \dodoi{10.1088/2041-8205/783/2/L35}

\bibitem[{{Fan} {et~al.}(2002){Fan}, {Narayanan}, {Strauss}, {White}, {Becker},
  {Pentericci}, \& {Rix}}]{2002AJ....123.1247F}
{Fan}, X., {Narayanan}, V.~K., {Strauss}, M.~A., {et~al.} 2002, \aj, 123, 1247,
  \dodoi{10.1086/339030}

\bibitem[{{Fender} {et~al.}(2015){Fender}, {Stewart}, {Macquart}, {Donnarumma},
  {Murphy}, {Deller}, {Paragi}, \& {Chatterjee}}]{2015aska.confE..51F}
{Fender}, R., {Stewart}, A., {Macquart}, J.~P., {et~al.} 2015, in Advancing
  Astrophysics with the Square Kilometre Array (AASKA14), 51,
  \dodoi{10.22323/1.215.0051}

\bibitem[{{Fialkov} \& {Loeb}(2017)}]{2017ApJ...846L..27F}
{Fialkov}, A., \& {Loeb}, A. 2017, \apjl, 846, L27,
  \dodoi{10.3847/2041-8213/aa8905}

\bibitem[{{Fukugita} {et~al.}(1998){Fukugita}, {Hogan}, \&
  {Peebles}}]{1998ApJ...503..518F}
{Fukugita}, M., {Hogan}, C.~J., \& {Peebles}, P.~J.~E. 1998, \apj, 503, 518,
  \dodoi{10.1086/306025}

\bibitem[{{Furlanetto} \& {Oh}(2008{\natexlab{a}})}]{2008ApJ...681....1F}
{Furlanetto}, S.~R., \& {Oh}, S.~P. 2008{\natexlab{a}}, \apj, 681, 1,
  \dodoi{10.1086/588546}

\bibitem[{{Furlanetto} \& {Oh}(2008{\natexlab{b}})}]{2008ApJ...682...14F}
---. 2008{\natexlab{b}}, \apj, 682, 14, \dodoi{10.1086/589613}

\bibitem[{{Hallinan} {et~al.}(2019){Hallinan}, {Ravi}, {Weinreb}, {Kocz},
  {Huang}, {Woody}, {Lamb}, {D'Addario}, {Catha}, {Law}, {Kulkarni}, {Phinney},
  {Eastwood}, {Bouman}, {McLaughlin}, {Ransom}, {Siemens}, {Cordes}, {Lynch},
  {Kaplan}, {Brazier}, {Bhatnagar}, {Myers}, {Walter}, \&
  {Gaensler}}]{2019BAAS...51g.255H}
{Hallinan}, G., {Ravi}, V., {Weinreb}, S., {et~al.} 2019, in Bulletin of the
  American Astronomical Society, Vol.~51, 255,
  \dodoi{10.48550/arXiv.1907.07648}

\bibitem[{{Hashimoto} {et~al.}(2020){Hashimoto}, {Goto}, {On}, {Lu}, {Santos},
  {Ho}, {Wang}, {Kim}, \& {Hsiao}}]{2020MNRAS.497.4107H}
{Hashimoto}, T., {Goto}, T., {On}, A. Y.~L., {et~al.} 2020, \mnras, 497, 4107,
  \dodoi{10.1093/mnras/staa2238}

\bibitem[{{Hashimoto} {et~al.}(2021){Hashimoto}, {Goto}, {Lu}, {On}, {Santos},
  {Kim}, {Eser}, {Ho}, {Hsiao}, \& {Lin}}]{2021MNRAS.502.2346H}
{Hashimoto}, T., {Goto}, T., {Lu}, T.-Y., {et~al.} 2021, \mnras, 502, 2346,
  \dodoi{10.1093/mnras/stab186}

\bibitem[{{Heimersheim} {et~al.}(2022){Heimersheim}, {Sartorio}, {Fialkov}, \&
  {Lorimer}}]{2022ApJ...933...57H}
{Heimersheim}, S., {Sartorio}, N.~S., {Fialkov}, A., \& {Lorimer}, D.~R. 2022,
  \apj, 933, 57, \dodoi{10.3847/1538-4357/ac70c9}

\bibitem[{{Inoue}(2004)}]{2004MNRAS.348..999I}
{Inoue}, S. 2004, \mnras, 348, 999, \dodoi{10.1111/j.1365-2966.2004.07359.x}

\bibitem[{{Ioka}(2003)}]{2003ApJ...598L..79I}
{Ioka}, K. 2003, \apjl, 598, L79, \dodoi{10.1086/380598}

\bibitem[{{James} {et~al.}(2019){James}, {Ekers}, {Macquart}, {Bannister}, \&
  {Shannon}}]{2019MNRAS.483.1342J}
{James}, C.~W., {Ekers}, R.~D., {Macquart}, J.~P., {Bannister}, K.~W., \&
  {Shannon}, R.~M. 2019, \mnras, 483, 1342, \dodoi{10.1093/mnras/sty3031}

\bibitem[{{Jing} \& {Xia}(2022)}]{2022Univ....8..317J}
{Jing}, L., \& {Xia}, J.-Q. 2022, Universe, 8, 317,
  \dodoi{10.3390/universe8060317}

\bibitem[{{Kumar} \& {Linder}(2019)}]{2019PhRvD.100h3533K}
{Kumar}, P., \& {Linder}, E.~V. 2019, \prd, 100, 083533,
  \dodoi{10.1103/PhysRevD.100.083533}

\bibitem[{{Lau} {et~al.}(2021){Lau}, {Mitra}, {Shafiee}, \&
  {Smoot}}]{2021NewA...8901627L}
{Lau}, A. W.~K., {Mitra}, A., {Shafiee}, M., \& {Smoot}, G. 2021, \na, 89,
  101627, \dodoi{10.1016/j.newast.2021.101627}

\bibitem[{{Linder}(2020)}]{2020PhRvD.101j3019L}
{Linder}, E.~V. 2020, \prd, 101, 103019, \dodoi{10.1103/PhysRevD.101.103019}

\bibitem[{{Lorimer} {et~al.}(2007){Lorimer}, {Bailes}, {McLaughlin},
  {Narkevic}, \& {Crawford}}]{2007Sci...318..777L}
{Lorimer}, D.~R., {Bailes}, M., {McLaughlin}, M.~A., {Narkevic}, D.~J., \&
  {Crawford}, F. 2007, Science, 318, 777, \dodoi{10.1126/science.1147532}

\bibitem[{{Macquart} \& {Ekers}(2018)}]{2018MNRAS.480.4211M}
{Macquart}, J.~P., \& {Ekers}, R. 2018, \mnras, 480, 4211,
  \dodoi{10.1093/mnras/sty2083}

\bibitem[{{Macquart} {et~al.}(2010){Macquart}, {Bailes}, {Bhat}, {Bower},
  {Bunton}, {Chatterjee}, {Colegate}, {Cordes}, {D'Addario}, {Deller},
  {Dodson}, {Fender}, {Haines}, {Halll}, {Harris}, {Hotan}, {Johnston},
  {Jones}, {Keith}, {Koay}, {Lazio}, {Majid}, {Murphy}, {Navarro}, {Phillips},
  {Quinn}, {Preston}, {Stansby}, {Stairs}, {Stappers}, {Staveley-Smith},
  {Tingay}, {Thompson}, {van Straten}, {Wagstaff}, {Warren}, {Wayth}, {Wen}, \&
  {CRAFT Collaboration}}]{2010PASA...27..272M}
{Macquart}, J.-P., {Bailes}, M., {Bhat}, N.~D.~R., {et~al.} 2010, \pasa, 27,
  272, \dodoi{10.1071/AS09082}

\bibitem[{{Macquart} {et~al.}(2015){Macquart}, {Keane}, {Grainge}, {McQuinn},
  {Fender}, {Hessels}, {Deller}, {Bhat}, {Breton}, {Chatterjee}, {Law},
  {Lorimer}, {Ofek}, {Pietka}, {Spitler}, {Stappers}, \&
  {Trott}}]{2015aska.confE..55M}
{Macquart}, J.~P., {Keane}, E., {Grainge}, K., {et~al.} 2015, in Advancing
  Astrophysics with the Square Kilometre Array (AASKA14), 55,
  \dodoi{10.22323/1.215.0055}

\bibitem[{{Maity}(2024)}]{2024arXiv240805722M}
{Maity}, B. 2024, arXiv e-prints, arXiv:2408.05722,
  \dodoi{10.48550/arXiv.2408.05722}

\bibitem[{{McQuinn} {et~al.}(2009){McQuinn}, {Lidz}, {Zaldarriaga},
  {Hernquist}, {Hopkins}, {Dutta}, \&
  {Faucher-Gigu{\`e}re}}]{2009ApJ...694..842M}
{McQuinn}, M., {Lidz}, A., {Zaldarriaga}, M., {et~al.} 2009, \apj, 694, 842,
  \dodoi{10.1088/0004-637X/694/2/842}

\bibitem[{{Newburgh} {et~al.}(2016){Newburgh}, {Bandura}, {Bucher}, {Chang},
  {Chiang}, {Cliche}, {Dav{\'e}}, {Dobbs}, {Clarkson}, {Ganga}, {Gogo},
  {Gumba}, {Gupta}, {Hilton}, {Johnstone}, {Karastergiou}, {Kunz}, {Lokhorst},
  {Maartens}, {Macpherson}, {Mdlalose}, {Moodley}, {Ngwenya}, {Parra},
  {Peterson}, {Recnik}, {Saliwanchik}, {Santos}, {Sievers}, {Smirnov},
  {Stronkhorst}, {Taylor}, {Vanderlinde}, {Van Vuuren}, {Weltman}, \&
  {Witzemann}}]{2016SPIE.9906E..5XN}
{Newburgh}, L.~B., {Bandura}, K., {Bucher}, M.~A., {et~al.} 2016, in Society of
  Photo-Optical Instrumentation Engineers (SPIE) Conference Series, Vol. 9906,
  Ground-based and Airborne Telescopes VI, ed. H.~J. {Hall}, R.~{Gilmozzi}, \&
  H.~K. {Marshall}, 99065X, \dodoi{10.1117/12.2234286}

\bibitem[{{Pagano} \& {Fronenberg}(2021)}]{2021MNRAS.505.2195P}
{Pagano}, M., \& {Fronenberg}, H. 2021, \mnras, 505, 2195,
  \dodoi{10.1093/mnras/stab1438}

\bibitem[{{Petroff} {et~al.}(2019){Petroff}, {Hessels}, \&
  {Lorimer}}]{2019A&ARv..27....4P}
{Petroff}, E., {Hessels}, J.~W.~T., \& {Lorimer}, D.~R. 2019, \aapr, 27, 4,
  \dodoi{10.1007/s00159-019-0116-6}

\bibitem[{{Petroff} {et~al.}(2022){Petroff}, {Hessels}, \&
  {Lorimer}}]{2022A&ARv..30....2P}
---. 2022, \aapr, 30, 2, \dodoi{10.1007/s00159-022-00139-w}

\bibitem[{{Planck Collaboration} {et~al.}(2020)}]{2020AA...641A...6P}
{Planck Collaboration}, {et~al.} 2020, \aap, 641, A6,
  \dodoi{10.1051/0004-6361/201833910}

\bibitem[{{Prochaska} \& {Zheng}(2019)}]{2019MNRAS.485..648P}
{Prochaska}, J.~X., \& {Zheng}, Y. 2019, \mnras, 485, 648,
  \dodoi{10.1093/mnras/stz261}

\bibitem[{{Qiang} \& {Wei}(2021)}]{2021PhRvD.103h3536Q}
{Qiang}, D.-C., \& {Wei}, H. 2021, \prd, 103, 083536,
  \dodoi{10.1103/PhysRevD.103.083536}

\bibitem[{{Shannon} {et~al.}(2018){Shannon}, {Macquart}, {Bannister}, {Ekers},
  {James}, {Os{\l}owski}, {Qiu}, {Sammons}, {Hotan}, {Voronkov}, {Beresford},
  {Brothers}, {Brown}, {Bunton}, {Chippendale}, {Haskins}, {Leach},
  {Marquarding}, {McConnell}, {Pilawa}, {Sadler}, {Troup}, {Tuthill},
  {Whiting}, {Allison}, {Anderson}, {Bell}, {Collier}, {G{\"u}rkan}, {Heald},
  \& {Riseley}}]{2018Natur.562..386S}
{Shannon}, R.~M., {Macquart}, J.~P., {Bannister}, K.~W., {et~al.} 2018, \nat,
  562, 386, \dodoi{10.1038/s41586-018-0588-y}

\bibitem[{{Shaw} {et~al.}(2024){Shaw}, {Ghara}, {Beniamini}, {Zaroubi}, \&
  {Kumar}}]{2024arXiv240903255S}
{Shaw}, A.~K., {Ghara}, R., {Beniamini}, P., {Zaroubi}, S., \& {Kumar}, P.
  2024, arXiv e-prints, arXiv:2409.03255, \dodoi{10.48550/arXiv.2409.03255}

\bibitem[{{Singh} {et~al.}(2018){Singh}, {Subrahmanyan}, {Udaya Shankar},
  {Sathyanarayana Rao}, {Fialkov}, {Cohen}, {Barkana}, {Girish}, {Raghunathan},
  {Somashekar}, \& {Srivani}}]{2018ApJ...858...54S}
{Singh}, S., {Subrahmanyan}, R., {Udaya Shankar}, N., {et~al.} 2018, \apj, 858,
  54, \dodoi{10.3847/1538-4357/aabae1}

\bibitem[{{Sun} {et~al.}(2015){Sun}, {Zhang}, \& {Li}}]{2015ApJ...812...33S}
{Sun}, H., {Zhang}, B., \& {Li}, Z. 2015, \apj, 812, 33,
  \dodoi{10.1088/0004-637X/812/1/33}

\bibitem[{{Syphers} {et~al.}(2012){Syphers}, {Anderson}, {Zheng}, {Meiksin},
  {Schneider}, \& {York}}]{2012AJ....143..100S}
{Syphers}, D., {Anderson}, S.~F., {Zheng}, W., {et~al.} 2012, \aj, 143, 100,
  \dodoi{10.1088/0004-6256/143/4/100}

\bibitem[{{Syphers} {et~al.}(2009){Syphers}, {Anderson}, {Zheng}, {Haggard},
  {Meiksin}, {Chiu}, {Hogan}, {Schneider}, \& {York}}]{2009ApJ...690.1181S}
---. 2009, \apj, 690, 1181, \dodoi{10.1088/0004-637X/690/2/1181}

\bibitem[{{Thornton} {et~al.}(2013){Thornton}, {Stappers}, {Bailes},
  {Barsdell}, {Bates}, {Bhat}, {Burgay}, {Burke-Spolaor}, {Champion}, {Coster},
  {D'Amico}, {Jameson}, {Johnston}, {Keith}, {Kramer}, {Levin}, {Milia}, {Ng},
  {Possenti}, \& {van Straten}}]{2013Sci...341...53T}
{Thornton}, D., {Stappers}, B., {Bailes}, M., {et~al.} 2013, Science, 341, 53,
  \dodoi{10.1126/science.1236789}

\bibitem[{{Torchinsky} {et~al.}(2016){Torchinsky}, {Broderick}, {Gunst},
  {Faulkner}, \& {van Cappellen}}]{2016arXiv161000683T}
{Torchinsky}, S.~A., {Broderick}, J.~W., {Gunst}, A., {Faulkner}, A.~J., \&
  {van Cappellen}, W. 2016, arXiv e-prints, arXiv:1610.00683,
  \dodoi{10.48550/arXiv.1610.00683}

\bibitem[{{Vanderlinde} {et~al.}(2019){Vanderlinde}, {Liu}, {Gaensler}, {Bond},
  {Hinshaw}, {Ng}, {Chiang}, {Stairs}, {Brown}, {Sievers}, {Mena}, {Smith},
  {Bandura}, {Masui}, {Spekkens}, {Belostotski}, {Dobbs}, {Turok}, {Boyle},
  {Rupen}, {Landecker}, {Pen}, \& {Kaspi}}]{2019clrp.2020...28V}
{Vanderlinde}, K., {Liu}, A., {Gaensler}, B., {et~al.} 2019, in Canadian Long
  Range Plan for Astronomy and Astrophysics White Papers, Vol. 2020, 28,
  \dodoi{10.5281/zenodo.3765414}

\bibitem[{{Vedantham} {et~al.}(2016){Vedantham}, {Ravi}, {Hallinan}, \&
  {Shannon}}]{2016ApJ...830...75V}
{Vedantham}, H.~K., {Ravi}, V., {Hallinan}, G., \& {Shannon}, R.~M. 2016, \apj,
  830, 75, \dodoi{10.3847/0004-637X/830/2/75}

\bibitem[{{Virgili} {et~al.}(2011){Virgili}, {Zhang}, {O'Brien}, \&
  {Troja}}]{2011ApJ...727..109V}
{Virgili}, F.~J., {Zhang}, B., {O'Brien}, P., \& {Troja}, E. 2011, \apj, 727,
  109, \dodoi{10.1088/0004-637X/727/2/109}

\bibitem[{{Wanderman} \& {Piran}(2015)}]{2015MNRAS.448.3026W}
{Wanderman}, D., \& {Piran}, T. 2015, \mnras, 448, 3026,
  \dodoi{10.1093/mnras/stv123}

\bibitem[{{Xiao} {et~al.}(2021){Xiao}, {Wang}, \& {Dai}}]{2021SCPMA..6449501X}
{Xiao}, D., {Wang}, F., \& {Dai}, Z. 2021, Science China Physics, Mechanics,
  and Astronomy, 64, 249501, \dodoi{10.1007/s11433-020-1661-7}

\bibitem[{{Y{\"u}ksel} {et~al.}(2008){Y{\"u}ksel}, {Kistler}, {Beacom}, \&
  {Hopkins}}]{2008ApJ...683L...5Y}
{Y{\"u}ksel}, H., {Kistler}, M.~D., {Beacom}, J.~F., \& {Hopkins}, A.~M. 2008,
  \apjl, 683, L5, \dodoi{10.1086/591449}

\bibitem[{{Zaroubi}(2013)}]{2013ASSL..396...45Z}
{Zaroubi}, S. 2013, in Astrophysics and Space Science Library, Vol. 396, The
  First Galaxies, ed. T.~{Wiklind}, B.~{Mobasher}, \& V.~{Bromm}, 45,
  \dodoi{10.1007/978-3-642-32362-1_2}

\bibitem[{{Zhang}(2018)}]{2018ApJ...867L..21Z}
{Zhang}, B. 2018, \apjl, 867, L21, \dodoi{10.3847/2041-8213/aae8e3}

\bibitem[{{Zhang}(2023)}]{2023RvMP...95c5005Z}
---. 2023, Reviews of Modern Physics, 95, 035005,
  \dodoi{10.1103/RevModPhys.95.035005}

\bibitem[{{Zhang} {et~al.}(2023){Zhang}, {Zhao}, {Li}, {Zhang}, {Li}, \&
  {Zhang}}]{2023SCPMA..6620412Z}
{Zhang}, J.-G., {Zhao}, Z.-W., {Li}, Y., {et~al.} 2023, Science China Physics,
  Mechanics, and Astronomy, 66, 120412, \dodoi{10.1007/s11433-023-2212-9}

\bibitem[{{Zhang} {et~al.}(2021){Zhang}, {Zhang}, {Li}, \&
  {Lorimer}}]{2021MNRAS.501..157Z}
{Zhang}, R.~C., {Zhang}, B., {Li}, Y., \& {Lorimer}, D.~R. 2021, \mnras, 501,
  157, \dodoi{10.1093/mnras/staa3537}

\bibitem[{{Zheng} {et~al.}(2014){Zheng}, {Ofek}, {Kulkarni}, {Neill}, \&
  {Juric}}]{2014ApJ...797...71Z}
{Zheng}, Z., {Ofek}, E.~O., {Kulkarni}, S.~R., {Neill}, J.~D., \& {Juric}, M.
  2014, \apj, 797, 71, \dodoi{10.1088/0004-637X/797/1/71}

\end{thebibliography}
\bibliographystyle{aasjournal}

\end{document}